# Modeling of Pan Evaporation Based on the Development of Machine Learning Methods

Mustafa Al-Mukhtar [1*]

[1*]Civil Engineering Department. The University of Technology. Baghdad-Iraq;

mmalmukhtar@gmail.com

ORCID: 0000-0002-8850-0899

Abstract

For effective planning and management of water resources and implementation of the related strategies, it is important to ensure proper estimation of evaporation losses, especially in regions that are prone to drought. Changes in climatic factors, such as changes in temperature, wind speed, sunshine hours, humidity, and solar radiation can have a significant impact on the evaporation process. As such, evaporation is a highly non-linear, non-stationary process, and can be difficult to be modeled based on climatic factors, especially in different agro-climatic conditions. The aim of this study, therefore, is to investigate the feasibility of several machines learning (ML) models (conditional random forest regression, Multivariate Adaptive Regression Splines, Bagged Multivariate Adaptive Regression Splines, Model Tree M5, K- nearest neighbor, and the weighted K- nearest neighbor) for modeling the monthly pan evaporation estimation. This study proposes the development of newly explored ML models for modeling evaporation losses in three different locations over the Iraq region based on the available climatic data in such areas. The evaluation of the performance of the proposed model based on various evaluation criteria showed the capability of the proposed weighted K- nearest neighbor model in modeling the monthly evaporation losses in the studies areas with better accuracy when compared with the other existing models used as a benchmark in this study.

## 1. Introduction

The process of water leaving the surface of the earth and plant into the atmosphere is referred to as evapotranspiration (ET) (Jing et al. 2019; Tao et al. 2018). Hence, ET can be considered a combination of both biological and physical processes and this distinguishes it from evaporation which is completely a physical process (Malik et al. 2020; Yaseen et al. 2019). Evaporation (E) can be estimated via several methods, such as water balance, energy balance, mass transfer, Penman, and pan evaporation (PE) methods (Lundberg 1993; Zhao et al. 2013). However, the commonly used method of measuring evaporation is the pan evaporation method owing to its cost-effectiveness and ease of operation (Majhi et al. 2019; Wu et al. 2020). Globally, Class A pan is the adopted standard method of estimating water surface evaporation as it helps in comparatively studying the extent of evaporation in different regions (Masoner et al. 2008). But, the major drawback of its application in most developing countries is its costly nature (Ashrafzadeh et al. 2019). As per various studies, there is a possibility of converting the evaporation amount of both 60 cm and 20 cm diameter pans into that amount of Class A pan (Wu et al. 2020). Furthermore, the estimation of the evaporation level can be based on the available meteorological variables. Various models have been developed for evaporation estimation from metrological variables despite the non-linear relationship between evaporation and meteorological factors (Kisi et al. 2017). Some of these models are empirical such as Penman-Monteith; Priestley-Taylor; and Thornthwaite equations, and others are non-empirical such as catchment water budget approach; energy budget method; mass transfer method; and artificial intelligence methods.

One of the key hydrologic processes with a direct impact on the planning and operations of water resources is evaporation (Penman 1948; Stewart 1984). It is, therefore, essential that evaporation is accurately quantified by water resources managers and engineers (Qasem et al. 2019). Arid and semi-arid regions, such as in Iraq, normally witness high rates of E and this causes a significantly high level of water evaporation from water bodies, such as reservoirs (Sayl et al. 2016), river basins, and natural lakes into the atmosphere (Boers et al. 1986; Khan et al. 2019). Therefore, the extent of water loss from surface water bodies must be considered during the planning and operation of dams and other hydraulic structures for agricultural purposes and proper management of water resources (Moazenzadeh et al. 2018). The importance of evaporation in surface water balance is portrayed by the impact of climatic change (Sartori 2000). In recent years, the consequences of global warming on evaporative losses have increased the relevance of evaporation in water resources management (Eames et al. 1997; Priestley and Taylor 1972).

Evaporation can be estimated using both direct or indirect techniques (Moran et al. 1996; Penman 1948); it can be measured directly by estimating the rate of evaporation from a 1.22 m diameter $\times$ 0.25 m depth pan (the Class A pan) from a height of 0.15 m above the soil surface (Stanhill 2002). It has been proven that this approach gives an accurate estimate of evaporation rates over time. This approach is attractive due to its cost-effectiveness and ease of operation since it requires no form of installation of expensive pans and meteorological stations (Ghorbani et al. 2018). However, a major restriction of the Class A pan approach is that it cannot be implemented in different climate regions owing to its technical requirements. Therefore, studies have suggested the development of the indirect methods of evaporation estimation (especially the empirical and semi-empirical models) that are based on the use of various metrological parameters, such as wind speed (WS), sunshine hours (Sh), rainfall (RF), minimum temperature ($T_{min}$), relative humidity (RH), maximum temperature ($T_{max}$), and mean temperature ($T_{mean}$) (Ghorbani et al. 2018; Lu et al. 2018a). However, this form of estimation is mainly prone to certain drawbacks due to the non-linearity, non-stationary, and stochastic features of the meteorological variables used in building such models (Salih et al. 2019). This has necessitated the development of robust and reliable intelligent models for evaporation prediction and this has been the focus of numerous studies in the field of water engineering and resources management (Khan et al. 2018; Naganna et al. 2019).

The development of machine learning (ML) models has been remarkably progressed over the past two decades in different hydrological and climatological subjects. Such as rainfall (Salih et al. 2020; Yaseen et al. 2017), streamflow (Al-Sudani et al. 2019; Feng et al. 2020), drought (Malik et al. 2020; Mokhtarzad et al. 2017), surface water quality (Rezaie-Balf et al. 2020; Tao et al. 2018), geo-science (Maroufpoor et al. 2019; Sanikhani et al. 2018) and several others (Ali and Prasad 2019; Ali et al. 2020; Prasad et al. 2020). This is owing to the capacity of the ML models in solving complex problems associated with highly stochastic features (Chia et al. 2020). However, the implementation of ML models for modeling the evaporation process has been investigated in different regions using deferent models (e.g., artificial neural network (ANN), random forest (RF), support vector machine (SVM), gene expression programming (GEP), adaptive neuro-fuzzy inference system (ANFIS)) (Abghari et al. 2012; Baydaroğlu and Koçak 2014; Chia et al, 2020; Di et al. 2019; Fallah-Mehdipour et al. 2013; Fotovatikhah et al. 2018; Lu et al. 2018b, 2018a; Majhi et al. 2019; Moazenzadeh et al. 2018; Tabari et al. 2010). These ML models and their advanced versions have so far performed excellently in achieving good prediction accuracy (Falamarzi et al. 2014; Ghorbani et al. 2017; Yaseen et al. 2018). The other forms of ML models for evaporation estimation are always emphasized to be explored and investigated (Danandeh Mehr et al. 2018; Fahimi et al. 2016; Jing et al. 2019; Yaseen et al. 2018). Several studies have argued the generalization of the capability of the ML models over various climatic regions because it is believed that each climatic environment is associated with specific stochasticity and non-stationarity characteristics (Al-Mukhtar 2019).

As such, this study aims to evaluate the applicability and predictability of six different AI methods as a predictive model for the pan evaporation rates in Iraq. These methods were: Conditional random forest regression (Cforest), Multivariate Adaptive Regression Splines (MARS), Bagged Multivariate Adaptive Regression Splines (BaggedMARS), Model Tree M5, K- nearest neighbor (KNN), and the weighted k-nearest neighbor (KKNN). The 5 fold cross-validation was employed during the model's configuration so that the uncertainty inherited in models development can be reduced. The input parameters (predictors) were selected based on the Pearson correlation coefficient at a significance level of 0.05 and the Goodman-Kruskal tau measure. To the best of the knowledge, no previous works were performed to model the evaporation rates using these aforementioned models. As such, the novelty of the study can be demonstrated as being the first attempt to validate these methods in modeling such a highly complex relationship in arid and semi-arid areas of Iraq. All the modeling codes and plots visualization were written within the R version 3.6.1.

## 2. Case study and data explanation

In this study, data from three different agro-climatic regions in Iraq were selected to assess the performance of six AI methods (i.e. Cforest, MARS, BaggedMARS, M5, KNN, and KKNN) to model the monthly pan evaporation rates. The studied stations are located in Baghdad (33°20′26″N, 44°24′03″E, 41 m a.s.l), Basrah (30°31′58″N, 47°47′50″E, 6 m a.s.l), and Mosul (36°20′06″N, 43°07′08″E, 228 m a.s.l); which are located in the middle, south, and north of Iraq, respectively. The climate of the above regions is characterized as arid in Baghdad and Basrah to semi-arid in Mosul (Al-Mukhtar 2019).

The aforementioned predictive models were set up using various meteorological parameters i.e. maximum air temperatures (max. T), minimum air temperatures (min. T), average temperatures (T), wind speed (W), and relative humidity (RH) to predict the monthly pan evaporation (PE). The collected data extends throughout 1990–2013 for Baghdad and Mosul stations, and 1990–2012 for Basrah, based on a monthly time scale. The entire data were split into two datasets; 75% for calibration and 25% for validation. The cross-validation with 5-fold was employed during the training stage to prevent overfitting and to provide a tradeoff between model variability and bias. The data were standardized to be within 0 and 1 so that the small values have equal attention in the modeling procedure as the higher ones.

Table 1 presents the statistical description of the predictors and predictand parameters. In the table, the monthly statistical parameters maximum, minimum, mean, standard deviation, coefficient of variation, and skewness for the meteorological predictors (max. temperature, min. temperature, mean temperature, W, and RH) and predictand (PE) were calculated for the three stations. The maximum and minimum PE values were recorded in Basrah and Mosul, respectively, which might be attributed to the features of climatic zones i.e. Mosul is colder than Basrah. Furthermore, it is evident from the table that the maximum mean, standard deviation, positive skewness, and coefficient of variation of PE were also seen in Mosul and the minimum ones were in Basrah.

The correlation matrices of the meteorological variables in Baghdad, Basrah, and Mosul stations were depicted as shown in Figs 1,2, and 3, respectively. In these figures, the diagonal, top of the diagonal, and bottom of the diagonal describe the marginal distribution, bivariate distribution, and bivariate correlations of the respective variables, respectively. It is noticed from Figs 1, 2, and 3 that the temperatures, relative humidity, and pan evaporation have a bimodal distribution, while the wind speed is almost symmetric and its distribution is unimodal. The scatter plots (bottom triangular in the correlation matrix) show that the data sets of maximum temperatures, minimum temperatures, average temperatures, and relative humidity are evenly distributed along the fitted line (red line) with the pan evaporation. The maximum value of pan evaporation was recorded in Baghdad and Basrah, while the lower value was in Mosul. These variations in pan evaporations rates correspond to relevant temperatures variations.

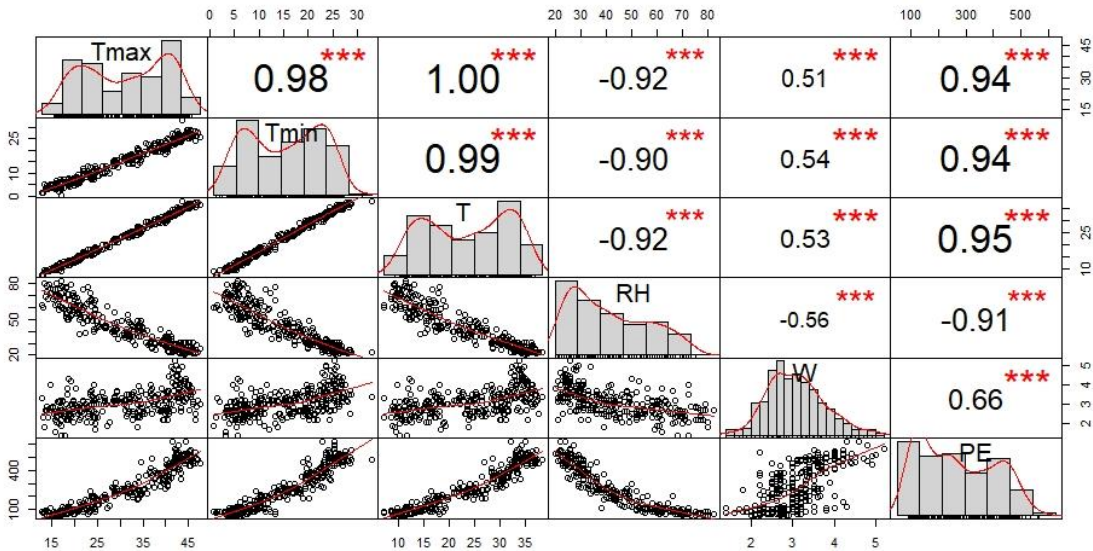

**Fig. 1** The statistical distribution, bivariate scatter, and correlations of the variables used at Baghdad station (red stars refer to the significant correlation)

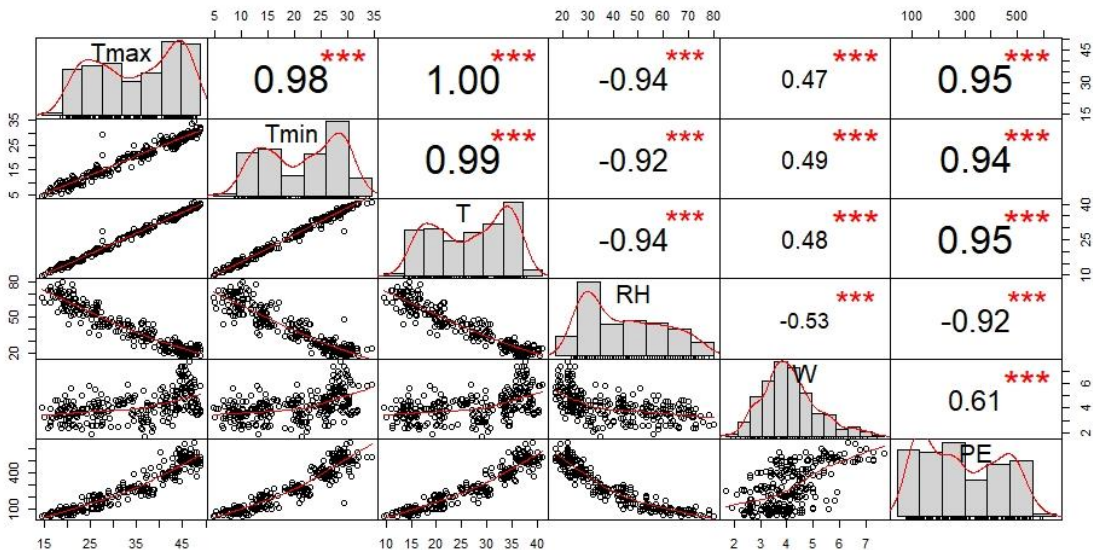

**Fig. 2** The statistical distribution, bivariate scatter, and correlations of the variables used at Basrah station (red stars refer to the significant correlation)

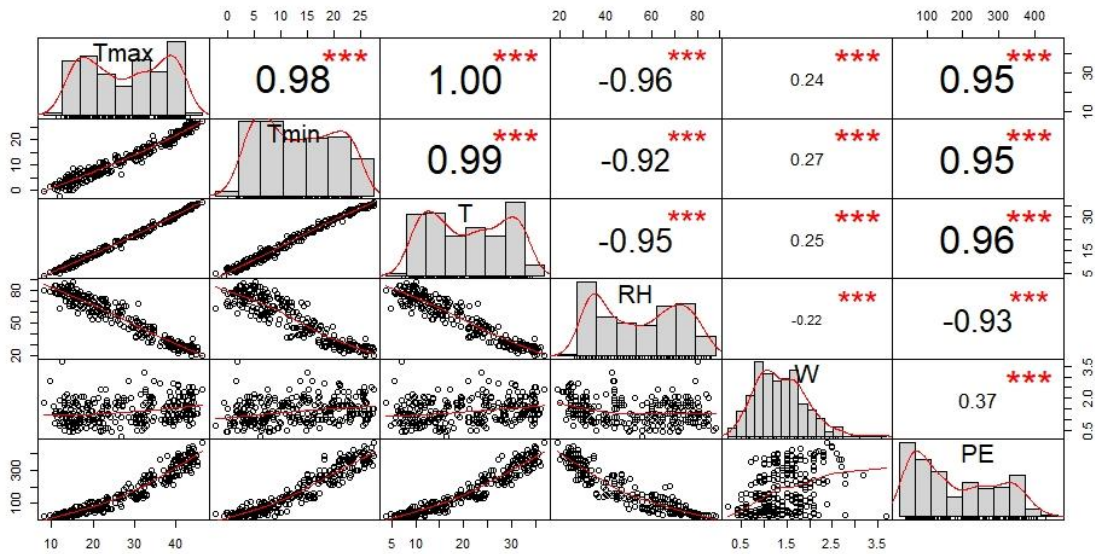

**Fig. 3** The statistical distribution, bivariate scatter, and correlations of the variables used at Mosul station (red stars refer to the significant correlation)

*2.1 Data Assurance Quality*

The data assurance quality is performed on any set of data acquired to provide better understanding and explore any problems in the data. At this point, the data is checked against any inconsistencies, missing values, outliers etc. Therefore, the meteorological data from the three studied stations were investigated to check if there is a problem with data quality. To this aim, dlookr package in R was employed to explore the data quality. Table 2 lists the diagnostic variables of data quality assurance used in this study. It is seen that there are no missing values in the entire datasets of the three stations. Additionally, the number of unique values and unique rate, which is unique count divided by number of observations, were also given in the table. It is evident that these values were not significantly varied from station to station and variable to another. The unique rates of the variables were almost identical across the three stations, indicating that the data are consistent.

Lastly, the zero values in addition to negative values, and number of outliers, which are useful measures to diagnose data integrity, were also determined (Table 2). It is evident that there are no zero and negative values recorded in the datasets. Moreover, the analysis revealed that there are no outliers in the meteorological parameters across the three stations except in wind speed. The number of detected outliers of wind speed in Baghdad, Basrah, and Mosul were only 4, 10, and 2 values, respectively, which however were not influential values given that the difference in mean of data with and without outliers was relatively small.

## 3. The developed Machine learning models

### 3.1 Conditional Random Forests

Random forest (Breiman 2001) has been widely used as a non-parametric regression tool in modeling many hydrological aspects. The conditional random forest (Cforest) is one variant of random forest methods (Tyralis et al. 2019). This method is a base learners-conditional inference trees which implements the random forest and bagging ensemble algorithms (Hothorn et al. 2006; Strobl et al. 2008; Strobl et al. 2009). On contrary to the random forest where the bootstrap aggregation is done by averaging predictions, the bagging scheme works by averaging observation weights extracted from each of the number of trees (Hothorn and Zeileis 2015). In the conditional inference tree, the variable importance selection is unbiased and the splitting induced by the recursive partition is not pruned to overfitting (Hothorn et al. 2006). In recursive partition, the hypothesis test for variable selection and stopping criteria are hypothesized which enhance its appropriateness to model any type of variable i.e. censored, ordinal, or multivariable response. The prediction accuracy of conditional inference tree is equivalent to the prediction accuracy of optimally pruned trees (Strobl et al. 2008). In addition, it is well known that the conditional inference tree can deal with large numbers of predictor variables even in the presence of complex interactions.

*3.2 Multivariate Adaptive Regression Splines (MARS)*

The MARS method is a non-parametric regression method introduced by Friedman (1991). The most featured characteristics of MARS are attributed to its capability on overcoming the limitations inherited in other regressions algorithms i.e. the global parametric modeling, non-parametric, and discontinuity (Friedman 1991). This approach is characterized as a combination of recursive partition with the spline fitting algorithms, making it a flexible regression modeling tool of high dimensional data (Kisi 2015). Recursive partitioning aims to adjust the coefficient values that best fit the data and to derive data-based basis functions (nodes) (Abraham et al. 2001). There are two ways to build the MARS model: forward and backward in which the recursive partitioning can be processed for selecting a subset of the regressor functions from the large entire basis. The basic idea of MARS is to approximate the function $f$ by several basis functions (such as low order polynomial functions $q$) over different sub-regions (knots) of the dependent-independent domain. The number of knots is controlling the tradeoff between the smoothness and accuracy of the function approximation given that the lowest order derivative is discontinues at every knot. So those, a set of basis function are constructed along the space of $q^{th}$ order splines approximation and optimizing the coefficient of these basis functions

The general form of the MARS model can be represented by:

$$\hat{f}(x) = a_o + \sum_{m=1}^{M} a_m \prod_{k=1}^{k_m} \left[ \delta_{km} \cdot (x_{v(k,m)} - t_{km}) \right]^1 +. \tag{1}$$

Where $a_o$ is a basis function constant, $M$ is the number of basis function, $\delta_{km} = \pm 1$, $k_m$ is the number of interactions of the basis function, $v(k,m)$ is a variable associated with m <sup>th</sup> basis function. $t_{km}$ is the knot location of the independent variable $x_{v(k,m)}$.

The generalized cross-validation criterion is employed to select the optimal model (Craven and Wahba 1979) (Kisi and Heddam 2019):

$$GCV(M) = \frac{1}{N} \sum_{i=1}^{N} \frac{\left[ y_i - \hat{f}(x_i) \right]^2}{\left[ 1 - \frac{C(M)}{N} \right]^2} \tag{2}$$

Where N is the number of data points, $y_i$ is the response variable, $C(M)$ is the penalty factor.

*3.3 Bagged Multivariate Adaptive Regression Splines (BaggedMARS)*

The BaggedMARS method is a Bootstrap Aggregating Multivariate Adaptive Regression Splines. In this approach, the resampling method (bagging) is employed to improve the prediction accuracy and stability. The method has been successfully applied in several hydrological studies (Chen et al. 2020; Chen et al. 2017). The bagging approach was produced by Breiman (1996) which is a method of making multiple releases of a predictor and uses these releases to get an aggregated value. The final aggregated value of the response variable can be either averaged overall outcomes (for regression), or plurality voted (for classification). The multiple releases are generated by bootstrap replicating of a learning set and employing these sets for further data generation. The main idea of the Bagged MARS is to build a MARS model for every bootstrap sample over the entire data space and then averaging these models (Equation 3)

$$\hat{f}_{baggedMARS} = \frac{1}{N}\sum_1^N \hat{f}(x) \tag{3}$$

*3.4 Model Tree M5*

M5 model tree is a decision tree system that constructs a tree-based piecewise linear model (Kisi et al. 2017; Quinlan 1992). The main concept of this method can be summarized as follows: Suppose we have a set of training cases where each set attributes to a specified target value (response value). Firstly, the model would calculate the standard deviation of the response variable of all set cases (T). Then, these T cases are split on the outcomes of the test. The potential outcomes are evaluated by determining the $T_i$ cases associated with each outcome. As a result, the expected reduction error is calculated by Equation 4 (Wang and Witten 1997):

$$\Delta_{error} = sd(T) - \sum_i \frac{|T_i|}{|T|} \times sd(T_i) \tag{4}$$

Where T is a set of cases in the data that reach a node, $sd$ is the standard deviation, $T_i$ is the subset of cases that have the i[th] outcomes of the potential test.

The M5 model relates the target values of the training sets to values of other attributes by multivariate linear model for the cases at every node using the standard regression method. Then, the surplus parameters induced from the above-constructed model are eliminated to reduce the estimated error.

Later, the pruning process is initiated by examining the non-leaf nodes starting from the root nodes. The final model selection is concluded based on the lower estimated error from either the above constructed linear model or the model subtree. At this point, if the linear model is chosen, the subtree at this node is considered (pruned) as a leaf. Lastly, the prediction accuracy is improved by the smoothing process where the predicted value at the leaf is adjusted using the following equation:

$$PV(S) = \frac{n_i \times PV(S_i) + k \times M(S)}{n_i + k} \tag{5}$$

Where $PV(S_i)$ is the predicted value at $S_i$, $n_i$ number of training cases, $S_i$ is the case follow a branch of subtree $S$, $k$ is the smoothing factor, $M(S)$ is the value given by the model at $S$,

*3.5 K-Nearest Neighbor Regression (KNN)*

The KNN method is a non-parametric technique that has been widely used in the field of supervised learning classification and regression (Goyal et al. 2012). The fundamental concept of KNN regression is to estimate the probabilities densities and regression functions through weighted local averaged of the dependent function (Lall and Sharma 1996). That is to be attained along with an estimation of conditional probability based on the K-nearest neighbors of the conditional probability of the vector $x$ (Cunningham and Delany 2007; Lall and Sharma 1996; Mehrotra and Sharma 2006). The density function used by KNN is estimated by: (Silverman 1986)

$$f_{NN(x)} = \frac{k/n}{V_{k(x)}} = \frac{k/n}{C_d r_k^d(x)} \tag{6}$$

Where k is the number of nearest neighbors, d is the vector space dimensions, $C_d$ is the sphere unit volume in d dimensions, $r_{k(x)}$ is the Euclidean distance to the $K^{th}$ nearest point value, $V_{k(x)}$ is the sphere volume of a d-dimensional with radius $r_{k(x)}$.

The choices of k patterns in the observations are determined based on their likeness to the conditioning vector using the Euclidean distance (Shabani et al. 2020). In KNN all predictors variables are assumed to have the same importance in estimating the conditional probability.

$$\xi_{t,i} = \sqrt{\sum_1^m \{S_j\,(x_{j,i} - x_{j,t})\}^2} \qquad (7)$$

Where $x$ is a vector with $m$ predictors $x_{j,i}$, $S_j$ is the scaling weight factor for the $j^{th}$ predictor.

Once the Euclidean distance is estimated for each projected feature vector, they are sort ascendingly. Thence, a set of K-NN cases is selected so that an element of the set records at time t is associated with the closest historical state with the current vector (Goyal et al. 2012).

*3.6 Weighted K-Nearest Neighbor (KKNN)*

The KKNN method is an extended version of the aforementioned KNN. The method characterized on the basic method as a more flexible tool and more closeness to the Loess regression, local regression estimator, which is a nonparametric technique that fits a smooth curve through points in a scatter plot using locally weighted regression (Schliep and Hechenbichler 2016). In this method, a higher weighted factor is given to the new observation (y, x) rather than those far away from (y, x). So, instead of considering the influence of nearest neighbors on the prediction as to the same for each of these classes in KNN, the distances are transformed into similarity measures and used as weights in KKNN. The algorithm structure of the KKNN is given below, however, the reader is referred to Schliep and Hechenbichler (2016) for further details:

1. If L is learning set of observations xi with given class membership yi of new observation x; then

$$L = \{(y_i, x_i), i = 1, \ldots, n_L\} \tag{8}$$

2. The k + 1 nearest neighbors to x are estimated according to the distance function $d_{(x, x_i)}$ i.e. Minkowski distance (a generalization of both the Euclidean distance and the Manhattan distance).

3. The above k+1 neighbor is employed to normalize the k smallest distances using the following equation:

$$D_t = D_{(x, x_i)} = \frac{d(x, x_i)}{d(x, x_{k+1})} \tag{9}$$

4. Then the new normalized distance is transformed into weights $w_{(i)} = k(D_{(i)})$ by using kernel function K(.).

5. Lastly, choose the class that shows a weighted majority of the K- nearest neighbors that belong to class r:

$$\hat{f} = max_r(\sum_{i=1}^{k} w_{(i)} I_{(y_{(i)} = r)}) \tag{10}$$

## 4. Models' development

To identify the best inputs combination for predicting the monthly pan evaporation from the selected meteorological parameters, two statistical approaches were employed; i.e. the traditional Pearson correlation coefficient and the Goodman and Kruskal association measure (Goodman and Kruskal 1979). In contrast to the parametric Pearson correlation statistics, which measures the linear association between numerical variables, the Goodman and Kruskal is a non-parametric measure for the association between the predictor and predictands relationships based on ordinal level variables. It explains the percentage improvement in predictability of the dependent variable (pan evaporation) given the value of other meteorological variables (Goodman and Kruskal 1979). The statistic is well known as an asymmetrical measure, which means that the explained fraction of variability in the categorical variable y explained by x is unequal to that of x explained by y. The models were set up using their default values and the predictive models were run until the minimum discrepancies between the observed and modelled data were obtained. Table 3 lists the final values of the tuning parameters and their descriptions for the evaluated methods.

## 5. Model's evaluation criteria

Various statistical metrics were employed in this study to evaluate the performances of the predictive models. These metrics include determination coefficient $R^2$, Root Mean Square Error RMSE, Mean Absolute Error MAE, Nash-Sutcliff Efficiency NSE, and Percentage Bias PBias. The following equations were employed to calculate the respective statistical criteria:

1. The determination coefficient

$$\mathbf{R^2} = \left[ \frac{\sum_{i=1}^{n}(O_i - \bar{O})(P_i - \bar{P})}{\sqrt{\sum_{i=1}^{n}(O_i - \bar{O})^2}\sqrt{\sum_{i=1}^{n}(P_i - \bar{P})^2}} \right]^2 \qquad (11)$$

Where: $O_i$ is the actual value, $\bar{O}$ is the average actual value, $P_i$ is the predicted value and $\bar{P}$ is the average predicted value. The values of $R^2$ is ranged between 0 and 1. The closer value to 1, the better model performance is.

2. Root mean square error

$$\mathbf{RMSE} = \sqrt{\frac{\sum_{i=1}^{n}(O_i - P)^2}{n}} \qquad (12)$$

The values of RMSE is ranged between 0 and $\infty$. The closer value to zero, the better model performance is.

3. Mean absolute error

$$\boldsymbol{MAE} = \frac{1}{n}\sum_{i=1}^{n}|P_i - O_i| \qquad (13)$$

The values of $MAE$ are ranged between 0 and $+\infty$. The closer value to 0, the better model performance is.

4. Nasch-Sutcliff efficiency coefficient

$$\mathbf{NSE} = 1 - \frac{\sum_{1}^{n}(O_i - p_i)^2}{\sum_{1}^{n}(O_i - \bar{P})^2} \qquad (14)$$

The values of NSE is ranged between $-\infty$ and 1. The closer value to 1, the better model performance is.

5. Percent Bias

$$\boldsymbol{PBIAS} = 100 \times \frac{\sum_{1}^{n}(O_i - P_i)}{\sum_{1}^{n}P_i} \qquad (15)$$

The values of $PBIAS$ are ranged between -100 and +100. The closer value to 0, the better model performance is.

Also, two different fitness functions were employed to support a comprehensive evaluation of the model's performance. The first fitness factor (F1) is given in equation 16, which measures the models' performance robustness by combining the MAE, RMSE, and $R^2$. So that providing multi objectives assessment criteria in one value (Chia et al. 2021). On the other side, F2 (equation 17) is an indicator based on Taylor's skill score (Taylor 2001) which takes into consideration the normalized RMSE and $R^2$.

$$F_1 = (RMSE + MAE)(1 - R^2) \qquad (16)$$

$$F_2 = \frac{4(1+R)^4}{(\delta + \frac{1}{\delta})^2 (1 + Rmax)^4} \qquad (17)$$

Where:

R is the correlation coefficient; $\delta$ is the standard deviation of modeled to the observed pan evaporation; $Rmax$ is the maximum value of the correlation coefficient i.e. 1. The above two indicators were normalized using equation 18 to be within 0 and 1. The closest value of fitness functions to 1, the better performance of the model.

$$x_{norm} = \frac{x - x_{min}}{x_{max} - x_{min}} \qquad (18)$$

Where

$x_{norm}$ is the normalized value; $x$ is the raw value; $x_{min}$ and $x_{max}$ are the minimum and maximum fitness function values, respectively.

## 6. Results and discussion

In this study, the pan evaporation rates in three meteorological stations of Iraq (Baghdad, Basrah, and Mosul) were modeled using developed novel artificial intelligence methods (i.e. Cforest, MARS, BaggedMARS, M5, KNN, and KKNN). The inputs combinations of these models over the three stations were selected based on their Pearson correlation coefficients with the dependent variable (PE) (Figs. 1-3); in addition to the Goodman and Kruskal tau measure (Figs. 4-6). The selection of the evaluated methods was based on their superior performance in modeling hydrological and water resources issues. Therefore, the aim was also to underline the application of these methods in solving many ill-posed water resources problems in Iraq.

Figs. 1, 2, and 3 show the Pearson correlation coefficient between the monthly pan evaporation (predictands) and the predictors for Baghdad, Basrah, and Mosul stations, respectively. All the employed predictors were significantly correlated with the predictand at a significance level of 0.05 (indicated as red stars). Thence, it is evident that there is a significant association between the predictands and predictors indicating the reliability of employing these variables for building predictive PE models. The correlation coefficient values of Tmax-PE, Tmin-PE, T-PE, RH-PE, and W-PE in Baghdad_ Basrah_ Mosul were 0.94, 0.94, 0.95, (-) 0.91, and 0.66 _ 0.95, 0.94, 0.95, (-) 0.92, and 0.61_ 0.95, 0.95, 0.96, (-) 0.93, and 0.37, respectively. In other words, only the relative humidity exhibits an inverse relationship with PE. Relative humidity describes how much water vapor is existing in the air relative to the full saturation capacity, that's to say that relative humidity is a measure of the atmosphere's capacity to hold water, which implies when it increases, the evaporation values decreases. From these figures, it is evident that the temperatures (maximum, minimum, and mean T C°), relative humidity %, wind speed (m/s) were all significantly associated at 0.05 level with the PE at the three stations. Justifying that the pan evaporation rates at these stations can be nicely modeled by these elements.

Figs. 4, 5, and 6 illustrate the results of Goodman and Kruskal tau measure for Baghdad, Basrah, and Mosul, respectively. In these figures, the diagonal elements illustrate the number of unique values for each variable (predictors and predictand), while the off-diagonal elements contain the forward and backward tau measures for each variable pair. That's to say, the numerical values in each element represent the association measure $\tau$ (x, y) from the variable x indicated in the row name to the variable y indicated in the column name. It can be raised from Fig. 4; the associations from Tmax, Tmin, T, RH, and W to pan evaporation in Baghdad station were 0.71, 0.62, 0.81, 0.24, and 0.17, respectively. In other words, all predictors were associated with the predictand values and hence providing evidence of potential predictability from the selected parameters to the monthly pan evaporation. In the same manner, the association from the predictors i.e. Tmax, Tmin, T, RH, and W to monthly pan evaporation in Basrah station were 0.70, 0.63, 0.82, 0.24, and 0.28, respectively (Fig. 5). While those in Mosul station were 0.72, 0.60, 0.84, 0.24, and 0.12, respectively (Fig. 6). Therefore, and based on the above analysis, the predictive models of monthly pan evaporation across the three stations were set up considering these five input parameters.

**Fig. 4** Goodman and Kruskal tau measure of association between predictors-predictand variables of Baghdad station

**Fig. 5** Goodman and Kruskal tau measure of association between predictors-predictand variables of Basrah station

**Fig. 6** Goodman and Kruskal tau measure of association between predictors-predictand variables of Mosul station

The obtained results from the employed models over the three stations during the training and validation were tabulated based on their statistical performances as shown in Table 4. The scatter plots between the observed and modeled monthly pan evaporation (mm) were depicted for Baghdad, Basrah, and Mosul as shown in Figs. 7, 8, and 9, respectively. Fig. 7 shows the scatter plot between observed and modeled monthly PE of the considered models during the validation stage at Baghdad station. It is obvious from the figure, that the closest scatter distribution around line 1:1 was obtained by KKNN and M5 model tree. That is to say that the KKNN and M5 methods outperformed the other evaluated methods to model the monthly pan evaporation. The statistical criteria: $R^2$,

RMSE, MAE, NSE, PBIAS obtained from the KKNN model during the validation stage were 0.98, 26.39, 18.62, 0.97, and 3.8, respectively. The performance of the M5 model tree was slightly less powerful than that of KKNN. Values of $R^2$, RMSE, MAE, NSE, PBIAS obtained from the M5 model during the validation stage were 0.97, 28.14, 20.69, 0.97, and -3.8. The MARS method performance was satisfactory but not as good as KKNN and M5. The statistical metrics: $R^2$, RMSE, MAE, NSE, PBIAS obtained from the MARS model during the validation stage were 0.97, 30.08, 22.39, 0.96, and 5.1, respectively, while those from BaggedMARS were 0.97, 30.74, 23.45, 0.96, and -5.6, respectively.

The performance of the Cforest method was not significantly different from those of MARS and BaggedMARS. The statistical metrics: $R^2$, RMSE, MAE, NSE, PBIAS obtained from the Cforest model during the validation stage were 0.96, 32.49, 23.05, 0.96, and -5, respectively. Lastly, the worst performance among the evaluated methods was obtained from KNN with statistical evaluation metrics $R^2$, RMSE, MAE, NSE, and PBIAS of 0.96, 35.81, 25.30, 0.95, and -6.7, respectively.

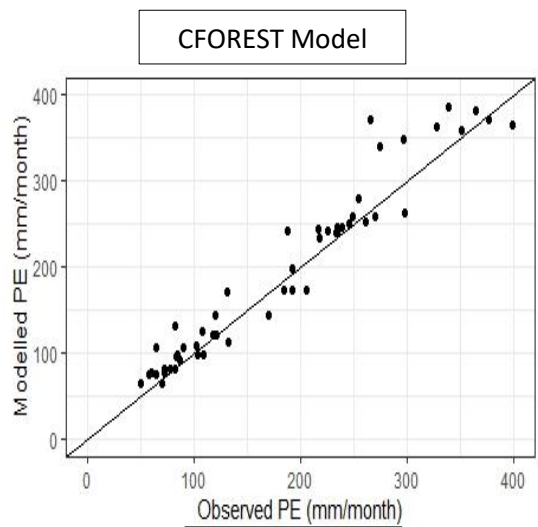

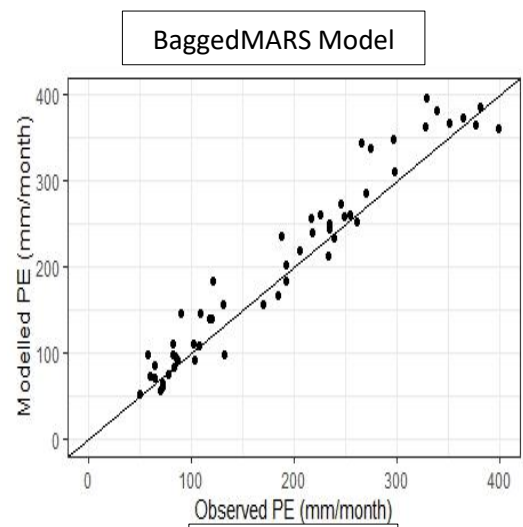

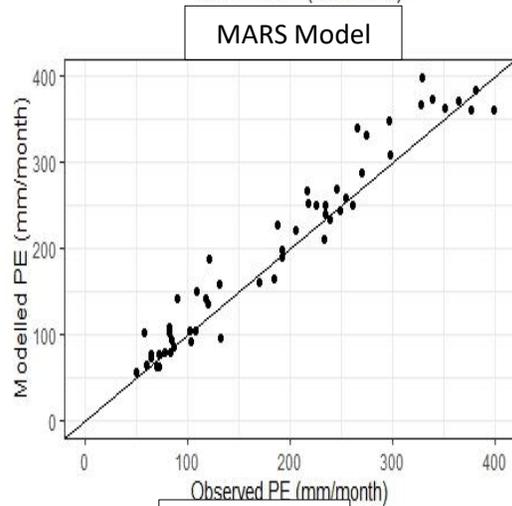

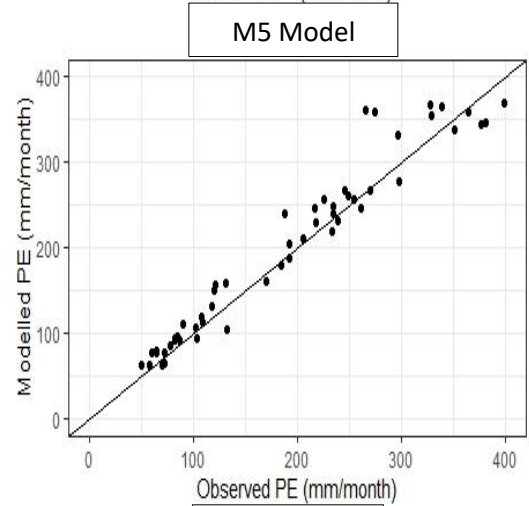

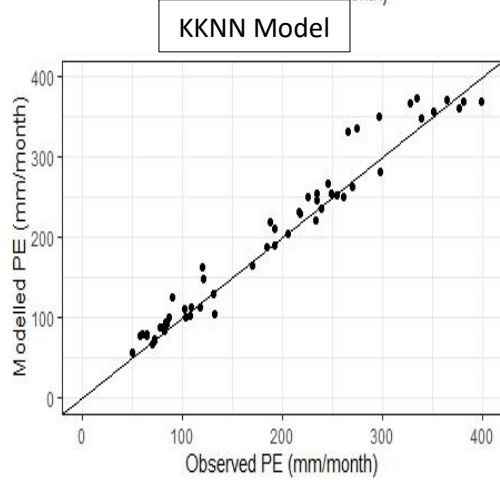

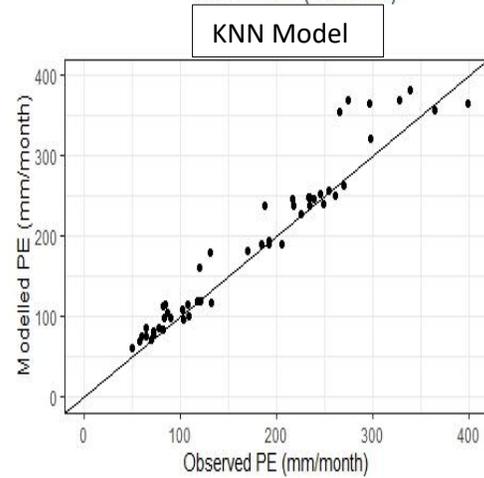

**Fig. 7** Scatter plots between observed and modelled monthly PE of the considered models during the validation stage at Baghdad station

Fig. 8 shows the scatter plot between observed and modeled monthly PE of the considered models during the validation stage at Basrah station. The KKNN and KNN results in addition to the M5 tree have the closest scattering around the line 1:1 in comparison to the other evaluated methods. The statistical metrics: $R^2$, RMSE, MAE, NSE, PBIAS obtained from the KKNN model during the validation stage were 0.97, 36.27, 27.82, 0.95, and -5.6, respectively. While those from KNN were 0.96, 34.63, 26.06, 0.96, and -5.1, respectively. The performance of the M5 model tree was almost akin to that of KKNN and KNN. Values of $R^2$, RMSE, MAE, NSE, PBIAS obtained from the M5 model during the validation stage were 0.96, 36.97, 27.11, 0.96, and -4.1, respectively. The monthly pan evaporation from the conditional Cforest method was sufficiently modeled though not as good as that of KKNN in terms of RMSE. Their statistical metrics values were 0.96, 38.22, 28.17, 0.95, and -5.1, respectively. Lastly, the lowest performances were obtained from the MARS and BaggedMARS. The performances criteria ($R^2$, RMSE, MAE, NSE, PBIAS) of MARS (BaggedMARS) were 0.95, 44.38, 36.79, 0.93, and -8.1, respectively (0.96, 41.45, 34.52, 0.94, and -7.1, respectively).

In Mosul station, the evaluated method's performances were quite similar to that in Baghdad station. That's to say, the KKNN and M5 models keep their superior performance in comparison to the remainder methods as shown in the scatter plots (Fig. 9). The statistical criteria: $R^2$, RMSE, MAE, NSE, PBIAS obtained from the KKNN model during the validation stage were 0.97, 21.44, 14.54, 0.97, and 0.1, respectively. While those from M5 model were 0.96, 25.17, 17.61, 0.96, and 0.9, respectively. The performances of the above two methods were superior in comparison to that in Baghdad in terms of the predictive capability which is represented by the lesser RMSE, MAE, and PBIAS (Table 4). The performance of the Cforest method was slightly different from those of KKNN and M5. The statistical metrics: $R^2$, RMSE, MAE, NSE, PBIAS obtained from the Cforest model during the validation stage were 0.96, 25.64, 18.43, 0.96, and -2.2, respectively.

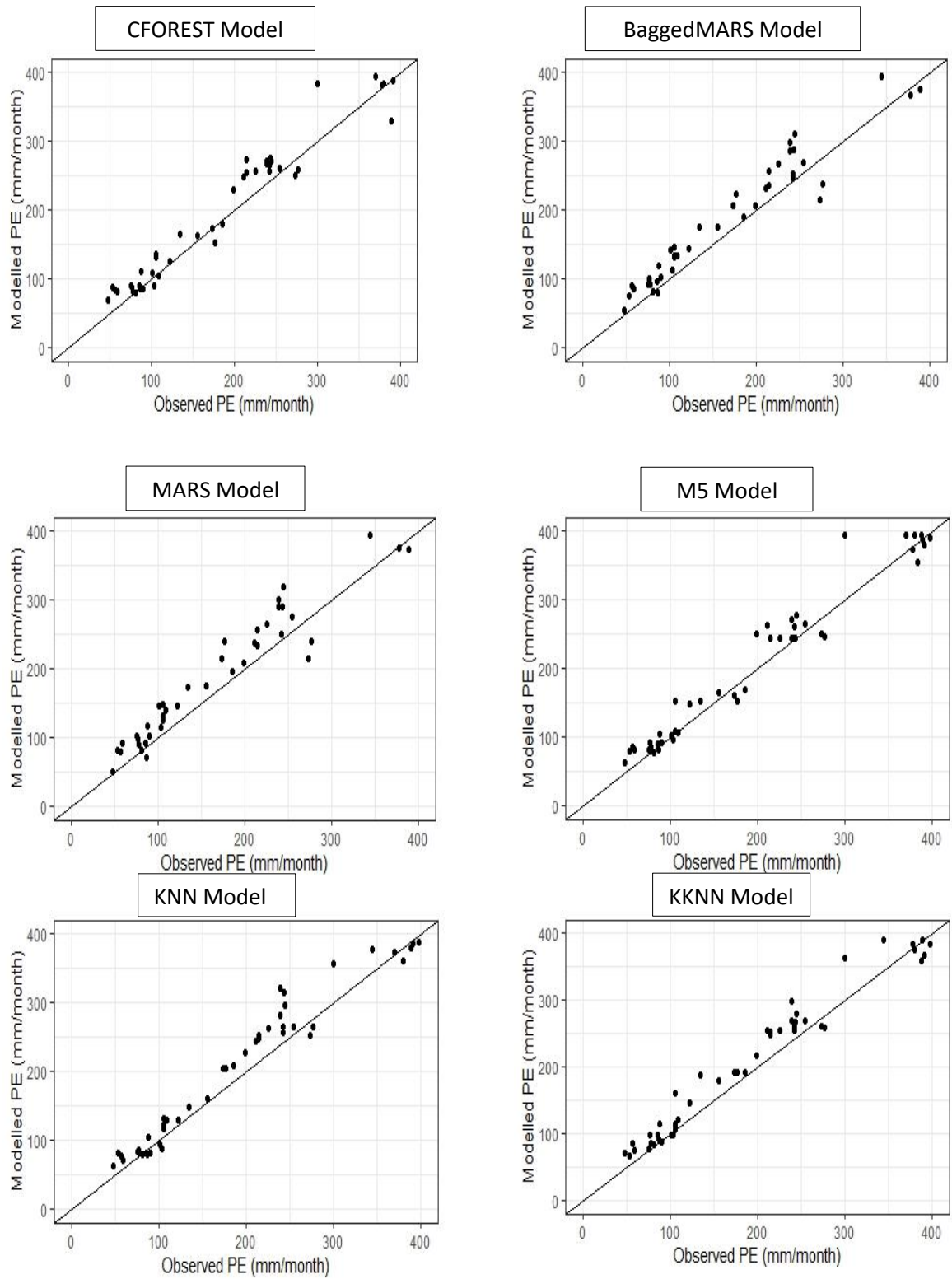

**Fig. 8** Scatter plots between observed and modelled monthly PE of the considered models during the validation stage at Basrah station

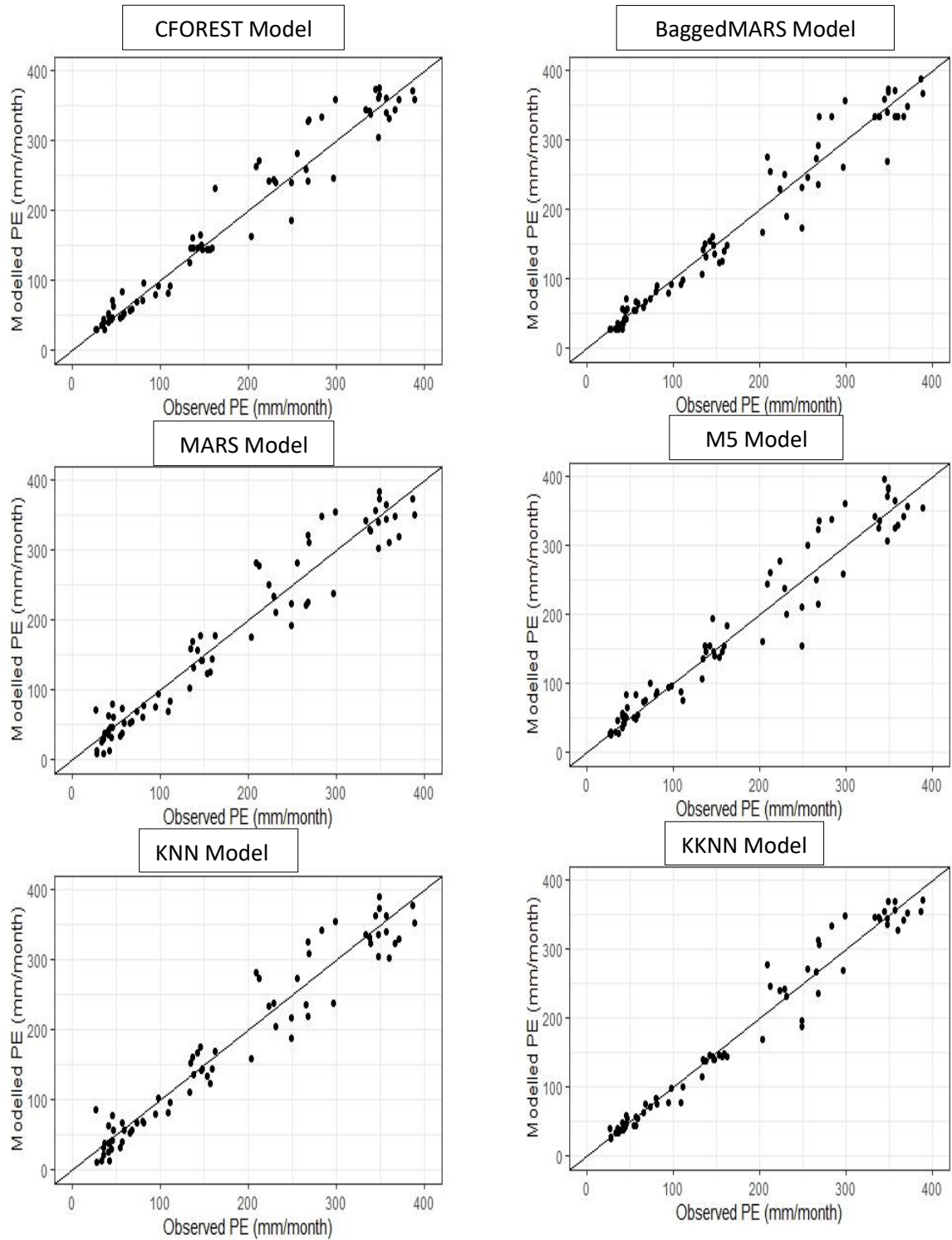

**Fig. 9** Scatter plots between observed and modelled monthly PE of the considered models during the validation stage at Mosul station

The predictive performance of the MARS and BaggedMARS methods were almost similar in terms of the employed statistical metrics values. The statistical metrics: $R^2$, RMSE, MAE, NSE, PBIAS obtained from the MARS (BaggedMARS) models during the validation stage were 0.94, 29.97, 23.73, 0.94, and 1.9 (0.94, 29.82, 23.87, 0.94, and 1.4), respectively. Lastly, the statistical evaluation metrics of KNN were 0.94, 29.61, 22.09, 0.94, and -2.3, respectively

It is proved from the above-presented results that the KKNN and M5 model tree were the optimal employed methods in Baghdad and Mosul. On the other side, In Basrah, the KNN slightly outperformed those optimal ones in Baghdad and Mosul. Arguably, the climatological driving forces of pan evaporation in Basrah are rather unlike those in other regions. As it is clear, Basrah is located to the south as a coastal area to the Arabian Gulf and this implies that it is easily exposed to the fluctuations of land-water interfaces arising from the energy exchanges. Basrah station recorded higher values of average monthly pan evaporation than Baghdad and Mosul which might be associated with the fact that this station is closer than the others to the equator (higher temperatures). All these factors exacerbate the non-linearity and stochastic complexity of the variable at the Basrah station.

The boxplot of the modeled monthly pan evaporation with their counterpart's values from the observations was plotted during the validation stage for Baghdad, Basrah, and Mosul as shown in Fig. 10. Where, the middle line represents the median values of monthly pan evaporation and the two whiskers refer to the maximum and minimum values of the modeled variable (PE) in the sample. While the $25^{th}$ and $75^{th}$ percentiles are represented by

24    the lower and upper edges of the box, respectively. In Baghdad, the box plot (Fig. 10a)

25    shows that the medians and percentiles of the modeled pan evaporation values from KKNN

26    and M5 were the closest to those of the observed values. The highest and lowest values

27    were also in good agreement with those of the observed data. The predictive capabilities

28    of the remaining methods were to a lesser extent than those of KKNN and M5.

29    Fig. 10b depicts the box plot at Basrah station, the KKNN, KNN, and M5 show

30    almost the optimal and consistent performance in terms of medians, quantiles, and

31    maximum/minimum values. The remaining predictive models (MARS, BMARS, and

32    CFR) were somehow unlike and inconsistent with the observed data set.

33    The optimal prediction performances of the evaluated models were obtained from

34    the data of Mosul station. The box plot (Fig. 10c) shows that the closest distribution of the

35    modeled medians and percentiles pan evaporation to those from the observed data were

36    from KKNN and M5. In other words, the performances here are just like that in Baghdad

37    but with higher predictive capacity. The highest and lowest values were perfectly matched

38    with those of the observed data.

39    The normalized fitness values calculated using F1, and F2 are tabulated in Table 5

40    for the three stations. As it can be drawn from the results in the table, the KKNN model

41    performed better than other models at the three stations with consistent F1 and F2 values

42    of 1. There is clear evidence that at different stations, the same model (KKNN) was suited

43    in estimating the monthly PE. This could be attributed to the characteristics of KKNN

44    which is an extended technique of KNN. In contrast to KNN, KKNN comprises a technique

45    that up to a certain degree is independent of a bad choice for k. In other words, the number

46  of nearest neighbors might be dominated by a small number of neighbors with large

47  weights whose classes influence the prediction because of their weights.

48      The other evaluated methods performed differently at each station. For example, at

49  Baghdad station, BaggedMARS and MARS performed better than KNN, Cforest, and M5

50  according to F1 and F2. While at Mosul station, Cforest was superior in comparison to the

51  KNN, BaggedMARS, MARS, and M5. The performance of the fitness function at Basrah

52  was not consistent when using F1 and F2. BaggedMARS has the highest F1 value (0.617)

53  and the M5 model attained the highest F2 value (0.333). The variation in the performance

54  of the fitness functions could be attributed to the unique characteristics of each one, which

55  in turn emphasizes the role of the fitness function in identifying the best performance

56  model. The good performance of KKNN was in agreement with findings of previous works

57  (e.g. Naghibi et al. 2020; Feng and Tian 2020) where they reported superior performances

58  of this model in comparison to other machine learning methods.

59      The successful applications of AI methods in this study were consistent with

60  elsewhere semi-arid case studies (Üneş et al. 2020; Zounemat-Kermani et al. 2019). For

61  instance, Ghorbani et al. (2018) used a hybrid predictive Multilayer Perceptron-Firefly

62  Algorithm for the prediction of daily pan of two station at Northern Iran using six different

63  combinations of the predictor variables, max T, min T, W, sun shine hours, and RH. They

64  compared their results with the standalone MLP method using several statistical metrics.

65  They concluded a superior performance of the applied methods in terms of the statisticial

66  indices used. Guan et al. (2020) applied both classical SVR models and SVR hybrid models

67  for prediction of daily pan evaporation across the three stations at coastal region of Iran.

68    They also employed the meteorological variables of max T, RH, precipitation, sun shine

69    hours, W, and radiation to set up the predictive models. They reported that the performance

70    of classical and hybrid models improved with increase in number of parameters so that the

71    inclusion of additional parameters results in a consistent improvement in the performance

72    of PE estimation model. However, in this study, the applicability of newly explored AI

73    methods were investigated on three meteorological stations in Iraq for predicting higher

74    evaporation rates using max. T, min. T, average temperatures (T), W, and RH.

75    One of the most advantageous applications of the KNN method is being not

76    sensitive to the outliers. It also imposes no force on the samples to satisfy specific

77    distribution. KKNN is simpler than tree-based methods in terms of running time and

78    produces acceptable outcomes. However, some drawbacks should be considered when

79    using this method i.e. in contrast to decision tree methods, it is not possible to impose rules

80    e.g. the number of classes or nodes (Shabani et al. 2020; Naghibi and Dashtpagerdi 2017).

81    However, it is noteworthy that the accuracy and performances of the applied

82    methods are not consistent in different climatic characteristics. So, for future work, it is

83    highly recommended to develop a machine learning method for homogeneous climatic

84    regions, which will be the subject for future work.

85

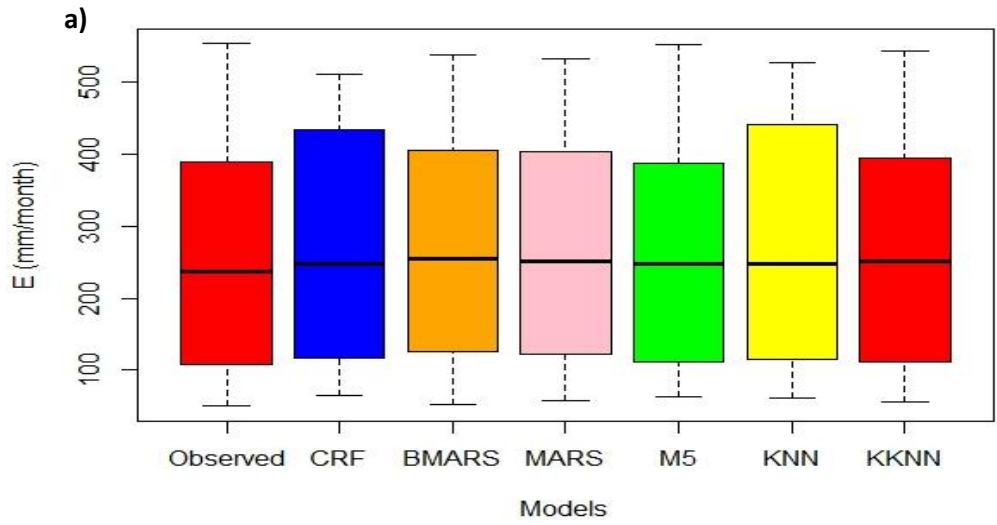

86

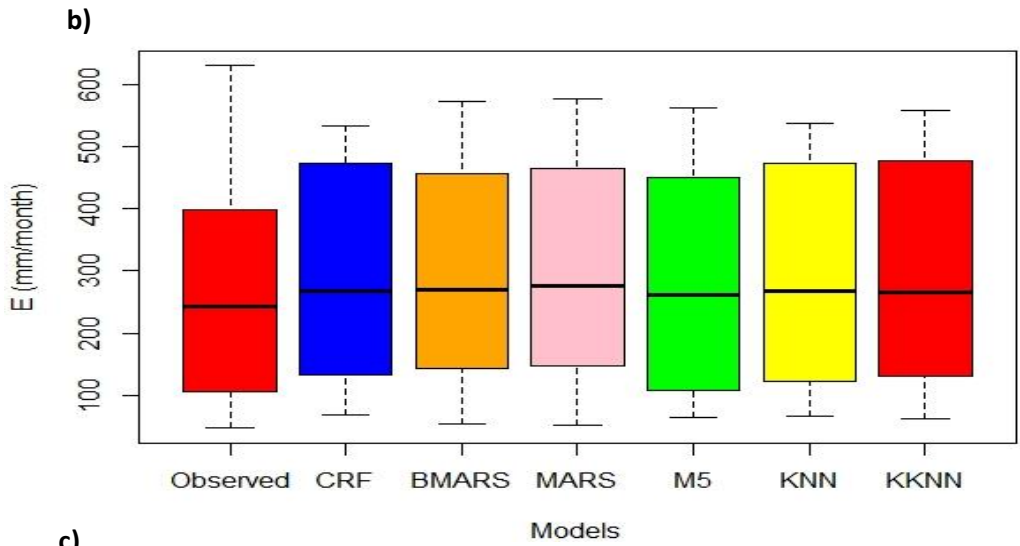

87

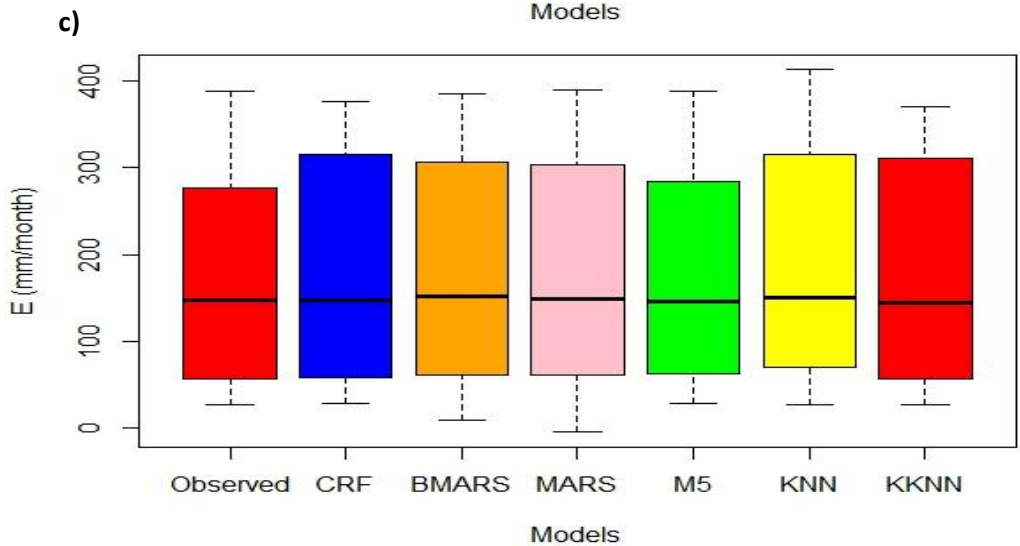

88

89          **Fig. 10** Box plot of the modelled and observed monthly pan evaporation at a)
90                    Baghdad station b) Basrah station and c) Mosul station

91     **7. Conclusions**

92     In this study, six models were employed to predict the monthly pan evaporation rates in

93     three different meteorological stations in Iraq (Baghdad, Basrah, and Mosul), and their

94     performances were evaluated against the observed data sets. The evaluated models

95     included: the conditional random forest regression, Multivariate Adaptive Regression

96     Splines, Bagged Multivariate Adaptive Regression Splines, Model Tree M5, K- nearest

97     neighbor, and the weighted K- nearest neighbor. The 5 fold cross-validation was adopted

98     during the model's configuration so that the uncertainty inherited in models developments

99     can be inhibited. The input parameters (predictors) were selected based on the Pearson

100    correlation and the Goodman and Kruskal tau measure. Therefore, temperatures

101    (maximum, minimum, and mean), relative humidity, and wind speed were chosen as they

102    show a significant association with the predictand values (PE). It was concluded that all

103    the evaluated methods satisfactory modeled the pan evaporation though some

104    discrepancies in models performances were noticed. However, it was proved that the

105    KKNN and the model tree M5 were the soundest in terms of their predictive capability to

106    model the pan evaporation rates. This study proved the efficiency of using machine-

107    learning methods in handling complex non-linear relationships in water resources and shed

108    the light on using such newly developed models for further evaluation of other parameters

109    estimation.

110

111

112     **Table 1** The statistical description of the meteorological parameters

| Parameter | Station | Min. | Max. | mean | St. Dev | CV | skewness |
|---|---|---|---|---|---|---|---|
| Max. T (C°) | Baghdad | 12.90 | 47.70 | 31.09 | 10.32 | 0.33 | -0.13 |
| | Basrah | 14.60 | 48.90 | 33.81 | 10.47 | 0.31 | -0.17 |
| | Mosul | 8.30 | 46.40 | 28.21 | 11.05 | 0.39 | -0.01 |
| Min. T (C°) | Baghdad | 0.70 | 32.90 | 15.68 | 8.05 | 0.51 | -0.06 |
| | Basrah | 4.70 | 34.60 | 19.78 | 8.09 | 0.41 | -0.17 |
| | Mosul | -2.20 | 27.40 | 13.20 | 8.21 | 0.62 | 0.11 |
| RH (%) | Baghdad | 20.00 | 82.00 | 43.93 | 16.88 | 0.38 | 0.49 |
| | Basrah | 17.00 | 80.00 | 40.50 | 17.34 | 0.43 | 0.45 |
| | Mosul | 19.00 | 88.00 | 51.51 | 20.33 | 0.39 | 0.07 |
| Wind speed (m/sec) | Baghdad | 1.40 | 5.20 | 3.09 | 0.68 | 0.22 | 0.44 |
| | Basrah | 1.70 | 7.70 | 4.14 | 1.11 | 0.27 | 0.67 |
| | Mosul | 0.20 | 3.70 | 1.41 | 0.56 | 0.40 | 0.58 |
| Pan evaporation (mm) | Baghdad | 48.70 | 624.80 | 264.25 | 161.71 | 0.61 | 0.32 |
| | Basrah | 41.40 | 645.90 | 286.75 | 169.75 | 0.59 | 0.22 |
| | Mosul | 21.50 | 464.10 | 177.77 | 127.14 | 0.72 | 0.42 |

113     Table 2 The data quality assurance of the meteorological parameters

| | Variable | missing count | missing percent | unique count | unique rate | zero | minus | outlier |
|---|---|---|---|---|---|---|---|---|
| **Baghdad** | Tmax | 0 | 0 | 204 | 0.71 | 0 | 0 | 0 |
| | Tmin | 0 | 0 | 180 | 0.63 | 0 | 0 | 0 |
| | T | 0 | 0 | 244 | 0.85 | 0 | 0 | 0 |
| | RH | 0 | 0 | 69 | 0.24 | 0 | 0 | 0 |
| | W | 0 | 0 | 51 | 0.18 | 0 | 0 | 4 |
| | PE | 0 | 0 | 280 | 0.97 | 0 | 0 | 0 |
| **Basrah** | Tmax | 0 | 0 | 194 | 0.70 | 0 | 0 | 0 |
| | Tmin | 0 | 0 | 174 | 0.63 | 0 | 0 | 0 |
| | T | 0 | 0 | 234 | 0.85 | 0 | 0 | 0 |
| | RH | 0 | 0 | 66 | 0.24 | 0 | 0 | 0 |
| | W | 0 | 0 | 79 | 0.29 | 0 | 0 | 10 |
| | PE | 0 | 0 | 247 | 0.90 | 0 | 0 | 0 |
| **Mosul** | Tmax | 0 | 0 | 207 | 0.72 | 0 | 0 | 0 |
| | Tmin | 0 | 0 | 173 | 0.60 | 0 | 0 | 0 |
| | T | 0 | 0 | 250 | 0.87 | 0 | 0 | 0 |
| | RH | 0 | 0 | 69 | 0.24 | 0 | 0 | 0 |
| | W | 0 | 0 | 39 | 0.14 | 0 | 0 | 2 |
| | PE | 0 | 0 | 280 | 0.97 | 0 | 0 | 0 |

114     *missing_count : number of missing values*
115     *missing_percent : percentage of missing values*
116     *unique_count : number of unique values*
117     *unique_rate : rate of unique value. unique_count / number of observation*

118      *zero : number of observations with a value of 0*
119      *minus : number of observations with negative numbers*
120      *outlier : number of outliers*
121      *types : the data type of the variables*
122

123

124

Table 3 The tuning parameters of the evaluated methods

| Method | Parameters | | Description | parameters values of | | |
|---|---|---|---|---|---|---|
| | | | | Baghdad | Basrah | Mosul |
| Cforest | mtry | | number of predictors randomly sampled as candidates at each node for random forest | 2 | 2 | 5 |
| MARS | degree | | an integer specifying maximum interaction degree | 1 | 1 | 1 |
| BaggedMARS | degree | | an integer specifying maximum interaction degree | 1 | 1 | 1 |
| M5 | | nprune | number of terms in the pruned model (including intercept) | 8 | 9 | 4 |
| | pruned | | an argument to set up the pruning process | yes | yes | yes |
| | | smoothed | an argument to set up the smooth process | no | no | no |
| | | rules | an argument to generate a decision list for regression problems using separate-and-conquer | yes | no | yes |
| KNN | k | | the number of nearest neighbors | 9 | 7 | 5 |
| KKNN | Kmax | | the maximum number of neighbors considered. | 9 | 9 | 9 |
| | | distance | distance between neighbors | 2 | 2 | 2 |
| | | Kernel | Kernel function used | optimal | optimal | optimal |

## Table 4 Models performance criteria

| | Model | Training | | | | | Validation | | | | |
|---|---|---|---|---|---|---|---|---|---|---|---|
| | | $R^2$ | RMSE | MAE | NSE | PBIAS | $R^2$ | RMSE | MAE | NSE | PBIAS |
| **Baghdad** | Cforest | 0.97 | 28.59 | 19.66 | 0.97 | -0.3 | 0.96 | 32.49 | 23.05 | 0.96 | -5 |
| | MARS | 0.97 | 29.39 | 21.56 | 0.97 | 0 | 0.97 | 30.08 | 22.39 | 0.96 | -5.1 |
| | BaggedMARS | 0.97 | 28.61 | 21.15 | 0.97 | -0.1 | 0.97 | 30.74 | 23.45 | 0.96 | -5.6 |
| | M5 | 0.97 | 25.87 | 19.37 | 0.97 | 0 | 0.97 | 28.14 | 20.69 | 0.97 | -3.8 |
| | KNN | 0.96 | 32.96 | 23.02 | 0.96 | -0.6 | 0.96 | 35.81 | 25.3 | 0.95 | -6.7 |
| | KKNN | 0.98 | 24.05 | 17.02 | 0.98 | -0.3 | 0.98 | 26.39 | 18.62 | 0.97 | -3.8 |
| **Basrah** | Cforest | 0.97 | 30.35 | 22.46 | 0.96 | -0.3 | 0.96 | 38.22 | 28.17 | 0.95 | -5.1 |
| | MARS | 0.96 | 34.83 | 26.99 | 0.96 | 0 | 0.95 | 44.38 | 36.79 | 0.93 | -8.1 |
| | BaggedMARS | 0.96 | 33.55 | 26.03 | 0.96 | 0.2 | 0.96 | 41.45 | 34.52 | 0.94 | -7.1 |
| | M5 | 0.97 | 30.72 | 22.44 | 0.97 | 0 | 0.96 | 36.97 | 27.11 | 0.96 | -4.1 |
| | KNN | 0.97 | 31.04 | 22.74 | 0.96 | -0.6 | 0.96 | 34.63 | 26.06 | 0.96 | -5.1 |
| | KKNN | 0.98 | 26.53 | 19.78 | 0.97 | -0.2 | 0.97 | 36.27 | 27.82 | 0.95 | -5.6 |
| **Mosul** | Cforest | 0.96 | 24.19 | 17.16 | 0.96 | -0.4 | 0.96 | 25.64 | 18.43 | 0.96 | -2.2 |
| | MARS | 0.95 | 27.19 | 20.5 | 0.95 | 0 | 0.94 | 29.97 | 23.73 | 0.94 | 1.9 |
| | BaggedMARS | 0.95 | 27.81 | 20.82 | 0.95 | 0.2 | 0.94 | 29.82 | 23.87 | 0.94 | 1.4 |
| | M5 | 0.96 | 23.83 | 16.78 | 0.96 | 0 | 0.96 | 25.17 | 17.61 | 0.96 | 0.9 |
| | KNN | 0.95 | 27.31 | 19.7 | 0.95 | 0.8 | 0.94 | 29.61 | 22.09 | 0.94 | -2.3 |
| | KKNN | 0.97 | 21.29 | 14.05 | 0.97 | 0.1 | 0.97 | 21.44 | 14.54 | 0.97 | 0.1 |

Table 5 Normalized fitness factor at the three studied stations

|  |  | F1 | F2 |
|---|---|---|---|
| Baghdad | Cforest | 0.000 | 0.154 |
|  | BaggedMARS | 0.485 | 0.615 |
|  | MARS | 0.418 | 0.615 |
|  | M5 | 0.277 | 0.692 |
|  | KNN | 0.162 | 0.000 |
|  | KKNN | **1.000** | **1.000** |
| Basrah | Cforest | 0.230 | 0.250 |
|  | BaggedMARS | 0.617 | 0.250 |
|  | MARS | 0.171 | 0.000 |
|  | M5 | 0.137 | 0.333 |
|  | KNN | 0.000 | 1.000 |
|  | KKNN | **1.000** | **0.917** |
| Mosul | Cforest | 0.711 | 0.552 |
|  | BaggedMARS | 0.098 | 0.034 |
|  | MARS | 0.099 | 0.000 |
|  | M5 | 0.615 | 0.586 |
|  | KNN | 0.000 | 0.069 |
|  | KKNN | **1.000** | **1.000** |

Bold values represent the optimal ones

**Acknowledgment**: The author is greatly appreciated the anonymous reviewers of this article for their constructive comments. The authors wish to express their gratitude to the Ministry of water resources for providing the data

## Conflict of Interest

Conflict Of Interest- None

## Declarations

- The author has no relevant financial or non-financial interests to disclose.
- The author has no conflicts of interest to declare that are relevant to the content of this article.
- The author certifies that he has no affiliations with or involvement in any organization or entity with any financial interest or non-financial interest in the subject matter or materials discussed in this manuscript.
- The author has no financial or proprietary interests in any material discussed in this article.
- Code availability: available on request.
- Availability of data and material (data transparency): The datasets used during the current study are available from the corresponding author on reasonable request.
- Ethics approval (include appropriate approvals or waivers): Not applicable.
- Consent to participate (include appropriate statements): Not applicable.
- Consent for publication (include appropriate statements): The author gives consent to the publication of all details of the manuscript including texts, figures, and tables.

## Author Contributions

Conceptualization, Mustafa Al-Mukhtar; methodology, Mustafa Al-Mukhtar; software, Mustafa Al-Mukhtar.; validation, Mustafa Al-Mukhtar; writing—original draft preparation, Mustafa Al-Mukhtar; writing—review and editing, Mustafa Al-Mukhtar; visualization Mustafa Al-Mukhtar; The author have read and agreed to the published version of the manuscript.